\documentclass[prb,twocolumn]{revtex4-1}

\usepackage{amssymb}
\usepackage{amsmath}
\usepackage{graphicx}

\begin{document}

\title{Magnetic behavior of nanoparticles in patterned thin films}
\author{J. Escrig, P. Landeros, J.C. Retamal, and D. Altbir}
\affiliation{Departamento de F\'{i}sica, Universidad de Santiago de Chile, Casilla 307,\\
Santiago2, Chile}
\author{J. d' Albuquerque e Castro}
\affiliation{Instituto de F\'{i}sica, Universidade Federal do Rio de Janeiro, Caixa\\
Postal 68.528, RJ, 21941-972, Brazil}

\begin{abstract}
The magnetic behavior of truncated conical nanoparticles in
patterned thin films is investigated as a function of their size and shape.
Using a scaling technique, phase diagrams giving the relative stability of
characteristic internal magnetic structures of the particles are obtained.
The role of the uniaxial anisotropy in determining the magnetic properties
of such systems is discussed, and a simple method for stablishing its
strength is proposed.
\end{abstract}

\maketitle

Nanolithography techniques combined with material deposition and
pattern transfer processes have made possible the production of regular
arrays of magnetic particles with individual linear dimensions in the
submicron range. A relatively high degree of precision has been achieved in
the growth process of these structures, allowing the control of both the
particle size and the spacing between them. The samples thus produced
exhibit rather interesting properties, and show great potential for
technological applications \cite{ross01a,chapman98}.

Electrodeposition, and evaporation followed by lift-off are the two methods
that have been most widely used in the preparation of this sort of arrays or
patterned thin films \cite{ross01a}. The first is suitable for producing
cylindrically shaped particles in which the ratio $R$ between their height $%
H $ and diameter $D$ of the bottom basis (aspect ratio) is bigger than 1. On
the other hand, evaporation and lift-off process is more suitable for the
production of periodic arrays covering large areas, which may reach several
square cm \cite{ross01a}. Taking under consideration its low cost of
production, this method is seen as quite convenient for commercial
applications. Particles produced by evaporation have the shape of truncated
cones \cite{ross01b,meier98}, which can be characterized by three
parameters, namely $H$, $D$, and the ratio $\zeta =(D-d)/2H$, where $d$ is
the diameter of the top basis.

It is generally accepted that particles of ferromagnetic materials with
linear dimensions of the order of $10^{2}$ nm or less are too small to
accommodate a domain wall. As a consequence, one could envisage the
production of arrays of single domain ferromagnetic particles, which could
be used in the design of spintronic devices or as magnetic media for high
density recording \cite{ross01a,chou98}. In recent years, several groups
have concentrated their attention on the magnetic behavior of nanosized
particles \cite{ross01b,grimsditch98,abraham01,ross99,cowburn99,lebib01}. It
has been found that the internal magnetic structure of sufficiently small
particles is indeed rather close to that of a ferromagnet monodomain. Larger
particles, however, may also exhibit vortex-like configurations, depending
on their size and shape. Clearly, the determination of the conditions under
which these distinct magnetic configurations or phases are found is of great
relevance for practical applications. However, from the experimental point
of view, this is not a simple task since transitions between magnetic
configurations as the shape or size of the particle changes are not sharp 
\cite{cowburn99,lebib01}. Hence, theoretical investigations of the magnetic
behavior of such nanoparticles become highly desirable.

In this paper, we focus our attention on patterned thin films produced by
evaporation followed by lift-off. The systems we have in mind are Ni films,
which have been extensively studied in recent years \cite
{ross01b,meier98,farhoud00,hwang01}. We aim at determining, for various
shapes and sizes of the particles, the configuration of lowest energy out of
three characteristic ones. These configurations are: (I) ferromagnetic with
the magnetization parallel to the cone basis (in-plane); (II) ferromagnetic
with the magnetization perpendicular to the cone basis (out-of-plane); (III)
vortex-state with the magnetic moments laying parallel to the cone basis. We
shall deal with the situation in which the distance $L$ between the
particles in arrays is large enough (i.e. $L>D$) for the interaction between
them to be safely neglected \cite{ross01b}. In such cases, the magnetic
structure of each particle is determined solely by the internal interactions
its magnetic moments are subjected to, namely the short-ranged exchange
interaction, the classical dipolar interaction (which is responsible for the
shape anisotropy), and the magneto-crystalline anisotropy. It is worth
mentioning that transmission electron microscopy has clearly established
that both Ni and Co particles in samples prepared by evaporation are
polycrystalline, with elongated columnar structures \cite{ross01b}.
Micromagnetic calculations of the remanence in those particles indicate that
the inclusion of an out-of-plane crystalline anisotropy, in addition to the
magnetocrystalline anisotropy, is necessary in order to reproduce the
experimental data \cite{ross01b}. The importance of the uniaxial anisotropy
in stabilizing configuration II is discussed towards the end of this article.

Thus, we consider truncated conical Ni nanoparticles consisting of $N$
magnetic moments occupying sites of an underlaying fcc structure. We assume
that the cone axis (growth direction) is in the (111) direction \cite
{ross01b}. For fixed $\zeta $, the phase diagram giving the relative
stability of the three configurations described above is obtained by
comparing the corresponding total energies at each point in the $D$x$H$
plane. For each configuration $\{\overrightarrow{\mu }_{i}\}$ of the
magnetic moments, it is given by

\begin{equation}
E_{tot}(\{\overrightarrow{\mu }_{i}\},\{J_{ij}\},K)=\frac{1}{2}\sum_{i\neq
j}[E_{ij}-J_{ij}\overrightarrow{\mu }_{i}\cdot \overrightarrow{\mu }%
_{j}]+U_{K},
\end{equation}

\noindent where $E_{ij}$ is the dipolar interaction between magnetic moments 
$\overrightarrow{\mu }_{i}$ and $\overrightarrow{\mu }_{j}$, $J_{ij}$ is the
exchange coupling, and $U_{K}$ is the crystalline uniaxial anisotropy term
given by $U_{K}=-K\sum_{i=1}cos^{2}\theta _{i}$. Here, $\theta _{i}$ is the
angle between $\overrightarrow{\mu }_{i}$ and the cone axis. For the systems
under consideration, we take the lattice parameter $a_{0}=3.52\ \AA $, $\mu
_{i}=\mu =0.62\mu _{B}$, and $J_{i,j}=J=0.9$ mRy for first nearest-neighbors 
\cite{rosengaard97} and zero otherwise. Estimates of the anisotropy constant 
$K$ fall in the range $1.8\times 10^{5}-2.6\times 10^{5}$ erg/cm$^{3}$
[3,11], so we take for the moment the value $2\times 10^{5}$ [3].

The calculation of the above expression is a well defined problem. However,
with the present day standard computational facilities, it becomes rather
time-consuming as soon as the number $N$ of magnetic moments assumes values
comparable to those in real systems, i.e. $10^{8}$ or even $10^{9}$. As a
consequence, the computational effort necessary for obtaining a phase
diagram in which particles in such size range are included turns out to be
prohibitively large. Nevertheless, it can be significantly reduced with the
use of a scaling technique recently proposed and used to study the magnetic
properties of cylindrically shaped particles \cite{prl02}. Such technique
involves two steps: (a) calculations are performed for systems with the
exchange interaction $J$ scaled down by a factor $x<1$; (b) the $H$ and $D$
axes in the resulting phase diagrams are scaled up a factor $(1/x)^{\eta }$,
where $\eta $ is a positive constant (see below).

In which follows, we first demonstrate that the scaling technique can be
also used to obtain the phase diagram of truncated conical particles, even
in the presence of a strong uniaxial anisotropy, and show how $\eta $ is
obtained. Having done this, we use such technique to obtain phase diagrams
for different values of $\zeta $. Results are compared to existing
experimental data. Finally, we comment on the effects of the uniaxial
anisotropy on the magnetic behavior of those particles.

Fig.(1) shows results for cones with $\zeta =0.3$, and scaling factors $x=$
0.010 (squares), 0.015 (circles), and 0.020 (triangles). Dashed line
corresponds to the maximum height $H_{max}=D/2\zeta $ one particle can reach
for each diameter $D$. As in the case of cylinders \cite{prl02}, the curves
in the three phase diagrams have similar shapes, with a triple point that
moves along a common line as $x$ is varied. From the plotting of the height $%
H_{t}$ at the triple point as a function of $x$, we find that these two
quantities are related by the equation $H_{t}=Ax^{\eta }$, with $A\approx
74.4$ nm and $\eta \approx 0.55$. The uncertainty in the values of both $A$
and $\eta $ is smaller than 2\%. Within the numerical accuracy of the
present calculations, the value of $\eta $ that we obtained in this case
coincides with that for cylindrically shaped particles \cite{prl02}. It
means that $\eta $ is either independent of $\zeta $ or has a rather weak
dependence on such parameter. This point has been confirmed by the
calculations we have performed for other values of $\zeta $. Having
determined the value of $\eta $, we then scale up the three diagrams by
dividing the $D$ and $H$ axes by the corresponding $x$ to power $\eta $.
Results are shown in Fig.(2), where all points fall on a single set of
lines, representing the phase diagram for the full strength of the exchange
interaction ($x=1$). We point out that the slope of the line separating
phases I and II gives the critical value ($R_{c}$) of the aspect ratio $%
R=H/D $, which depends on the strength of the uniaxial anisotropy. We shall
return to this point below.

\begin{figure}[h]
\begin{center}
\includegraphics[width=8cm]{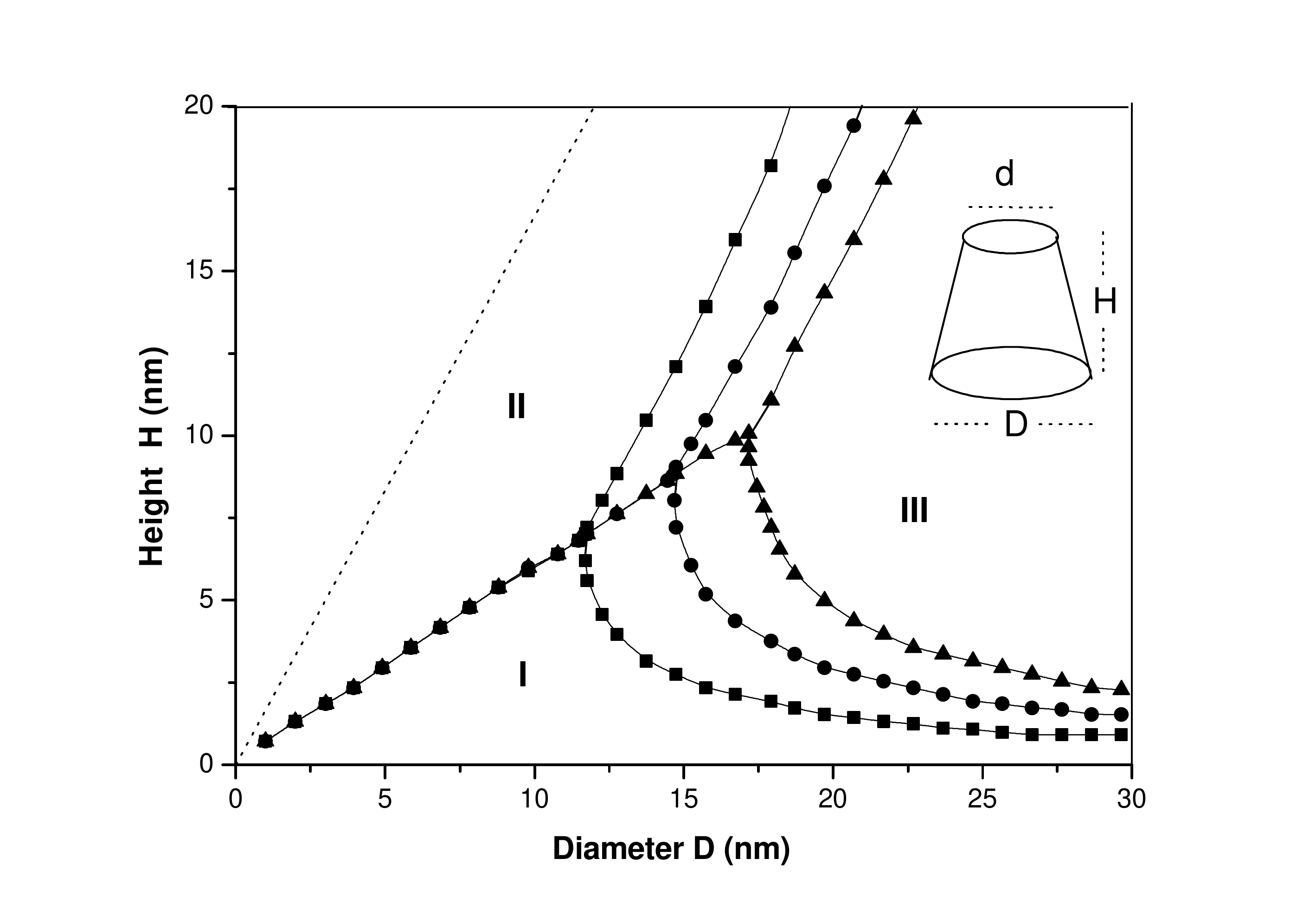}
\end{center}
\caption{Phase diagrams giving the relative stability of the in-plane
ferromagnetic (I), out-of-plane ferromagnetic (II), and vortex-like (III)
configurations of truncated conical Ni particles with $\zeta =0.3$, for
scaling factor $x=0.01$ (squares), 0.015 (circles) and 0.02 (triangles) .
Dashed line gives the maximum height $H_{max}$ of the particles as a
function of the diameter $D$ of their basis, for the specified value of $%
\zeta $. The inset illustrates the geometry of the particle under
consideration.}
\end{figure}

\begin{figure}[h]
\begin{center}
\includegraphics[width=8cm]{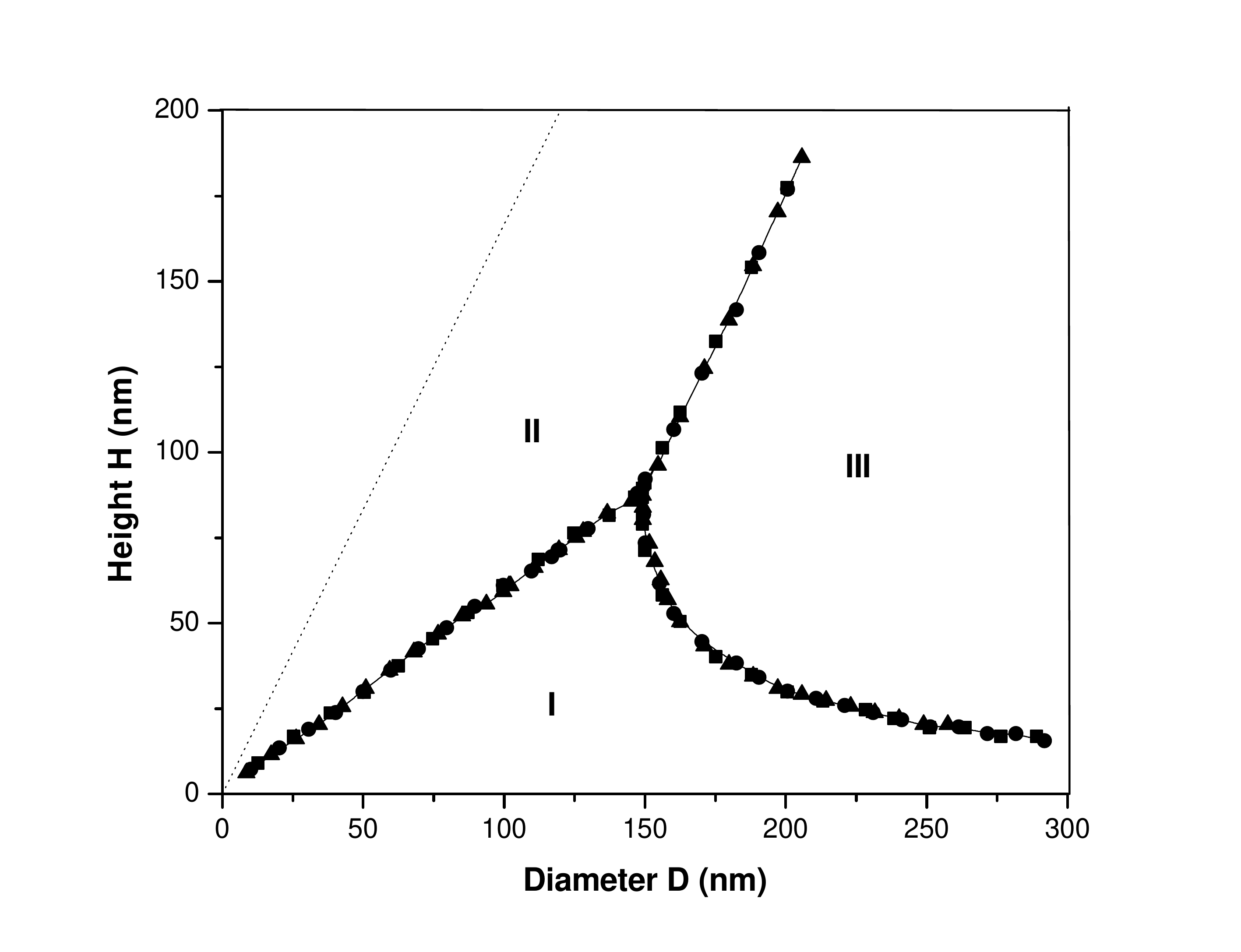}
\end{center}
\caption{Phase diagram obtained from those in Fig.(1), with the
corresponding axes scaled by a factor $1/x^{\eta }$, where $\eta =0.55$.
Dashed line corresponds to $H_{max}(D)$.}
\end{figure}

We are now in position to examine the phase diagrams for different values of 
$\zeta $ and full strength of the exchange coupling. Fig.(3) shows results
for $\zeta $ = 0 (squares), 0.2 (triangles), and 0.4 (circles). Dashed and
dotted lines correspond to $H_{max}(D)$ for $\zeta $ equals to 0.2 and 0.4,
respectively. We notice that as $\zeta $ increases from zero, that is to
say, as the shape of the particles deviates from that of a cylinder, changes
in the lines separating the three phases become less pronounced. In fact,
the lines for $\zeta $= 0.3 would appear in between those for 0.2 and 0.4,
and have been eliminated from the figure for the sake of clarity. These
results are confirmed by the experimental data of Ross {\it et al.}\cite
{ross01b} for Ni samples. These authors have investigated the magnetic
behavior of patterned thin films consisting of truncated conical particles
with $75\ \AA \leq D\leq 122\ \AA $, $0.43\leq R\leq 1.21$, and $0.14\leq
\zeta \leq 0.35$. They concluded that, independently of the value of $\zeta $%
, transitions from configuration I to II occurs at a critical aspect ratio $%
R_{c}=0.65$, which is in good agreement with the phase diagrams in Fig.(3).
Indeed, according to our results, for $\zeta >0$, transitions between these
two phases occur for $R_{c}\approx 0.61$.

\begin{figure}[h]
\begin{center}
\includegraphics[width=7cm]{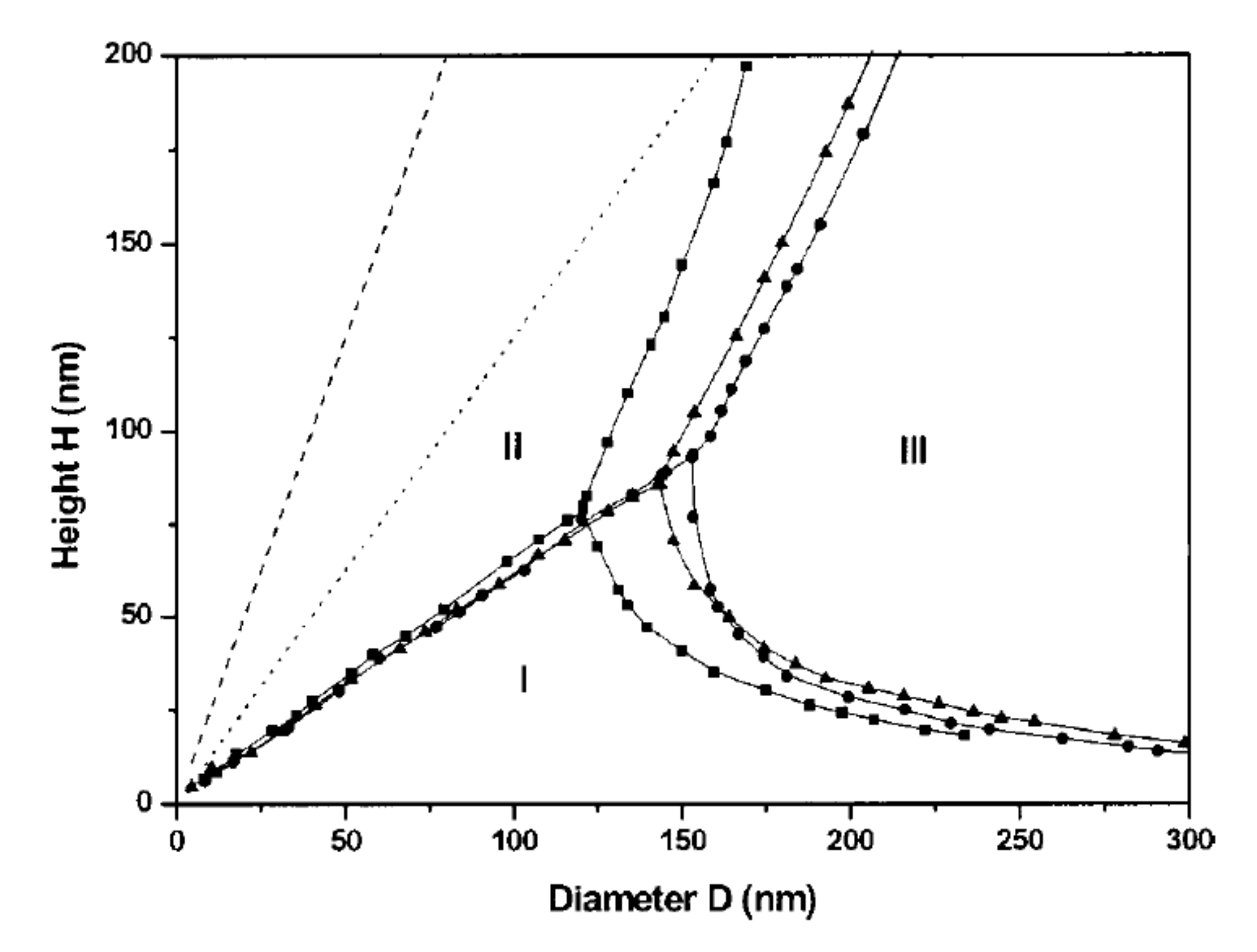}
\end{center}
\caption{Phase diagrams for cylindrically-shaped (squares) and cone-shaped
Ni particles with $\zeta =0.2$ (triangles) and 0.4 (circles). Dashed and
dotted lines correspond to $H_{max}(D)$ for $\zeta $ equals to 0.2 and 0.4,
respectively.}
\end{figure}

Although the boundaries between phases in the diagrams in Fig.(3) do not
change much as $\zeta $ is varied, the extent of region II (corresponding to
the out-of-plane configuration) is reduced as $\zeta $ increases. Such
region is bounded from above by the line $H_{max}=D/2\zeta $ , which means
that, for a fixed value of the uniaxial anisotropy constant $K$, a critical
value $\zeta _{c}=1/2R_{c}$ exists, above which phase II cannot be observed.
However, we expect the uniaxial anisotropy to favor the out-of-plane
configuration. Thus, $R_{c}$ should decrease as $K$ increases. This is
confirmed by the curves in Fig. (4), where $R_{c}$ is plotted as a function
of $K$ for different values of $\zeta $, namely zero (squares), 0.2
(triangles), and 0.4 (full circles). We find that for the whole range of
values of $\zeta $ considered here, the relation between $R_{c}$ and $K$ is
not much sensitive to that parameter. Indeed, for $\zeta \geq 0.2$ and $%
R_{c}<0.7$, the curves can be regarded as independent of $\zeta $. Apart
from confirming the intuitive idea that a strong uniaxial anisotropy is
necessary for stabilizing the out-of-plane configuration in particles with
large $\zeta $, Fig.(4) provides us with an extremely simple and practical
method for determining the value of $K$ in truncated conical particles. The
reason is the fact that the computational effort necessary for obtaining
such curve is minimum. In fact, since the border line between phases I and
II, for fixed values of $K$ and $\zeta $, is a straight line, its slope
(that is to say, $R_{c}$) can be obtained from calculations for small
particles. Thus, as a general procedure, the strength $K$ of the uniaxial
anisotropy necessary for obtaining the whole phase diagram can be
immediately determined from a similar graph and the measured value of
critical aspect ratio $R_{c}$, without much computational effort.

\begin{figure}[h]
\begin{center}
\includegraphics[width=8cm]{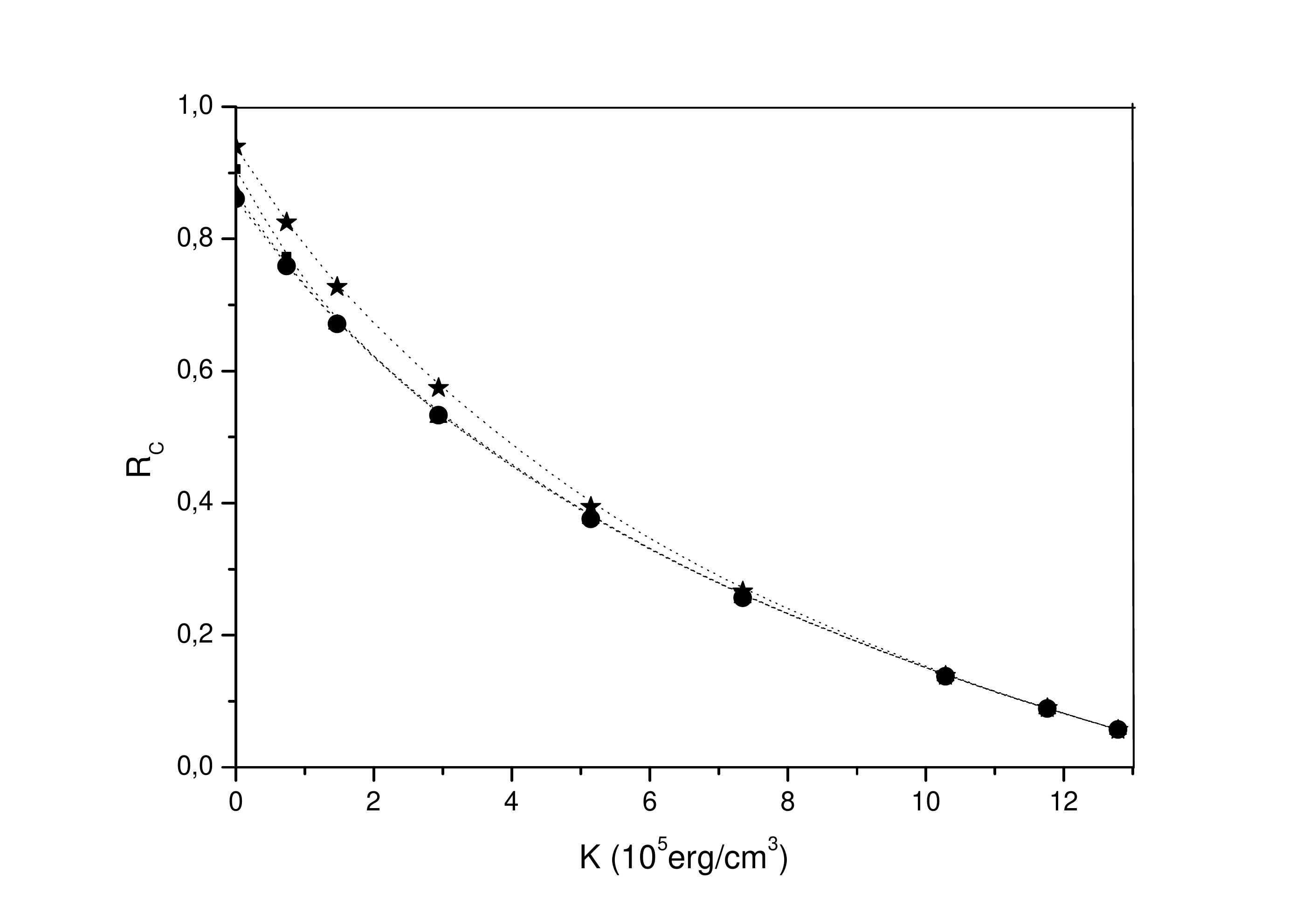}
\end{center}
\caption{The critical aspect ratio $R_{c}$ plotted as a function of the
uniaxial anisotropy constant $K$, for cylindrically-shaped (open squares)
and cone-shaped Ni particles with $\zeta =0.2$ (triangles) and 0.4 (circles).}
\end{figure}

In conclusion, we have investigated the magnetic behavior of truncated
conical particles, which exhibit a strong uniaxial anisotropy along the cone
axis. We have shown that even in such cases, the phase diagrams giving the
relative stability of three characteristic arrangements of the magnetic
moments within each particle can be obtained using a scaling method recently
proposed. We have found that the stability of the ferromagnetic out-of-plane
configuration strongly depends on the strength $K$ of the uniaxial
anisotropy, particularly when the shape of the particles deviates
significantly from that of a cylinder (i.e. for large $\zeta $). On the
basis of our results, we have proposed a simple method for determining the
value of $K$ from the measured critical aspect ratio $R_{c}$. We believe the
present work represents a relevant contribution to the understanding of the
magnetic properties of nanoparticles, and expect it to help the development
of new magnetic devices.

This work has been partially supported by Fondo Nacional de Investigaciones
Cient\'{i}ficas y Tecnol\'{o}gicas (FONDECYT), under Grants No. 1010127 and
7010127, and Grant P99-135F from the Millennium project of Chile, and by
CNPq, Rede Nacional de Nanoci\^{e}ncias/CNPq, FAPERJ, \ and \ the Millennium
Instituto de Nanoci\^{e}ncias of Brazil.

\end{document}